# Spin transport and accumulation in the persistent photoconductor Al$_{0.3}$Ga$_{0.7}$As

Jennifer Misuraca[1a], Joon-Il Kim[1], Jun Lu[2], Kangkang Meng[2], Lin Chen[2], Xuezhe Yu[2], Jianhua Zhao[2], Peng Xiong[1], and Stephan von Molnár[1]

[1] *Department of Physics, Florida State University, Tallahassee, Florida, 32306, United States*

[2] *State Key Laboratory of Superlattices and Microstructures, Institute of Semiconductors, Chinese Academy of Sciences, P.O. Box 912, Beijing 100083, China*

Electrical spin transport and accumulation have been measured in highly Si doped Al$_{0.3}$Ga$_{0.7}$As utilizing a lateral spin transport device. Persistent photoconductivity allows for the tuning of the effective carrier density of the channel material *in situ* via photodoping. Hanle effect measurements are completed at various carrier densities and the measurements yield spin lifetimes on the order of nanoseconds, an order of magnitude smaller than in bulk GaAs. These measurements illustrate that this methodology can be used to obtain a detailed description of how spin lifetimes depend on carrier density in semiconductors across the metal-insulator transition.

---

[a] Author to whom correspondence should be addressed. Electronic mail: jm05h@my.fsu.edu

All-electrical spin injection/detection and coherent spin transport/manipulation are necessary ingredients of any viable semiconductor spintronics technology. Electrical spin transport and accumulation have been achieved in several semiconducting materials in the last five years, including GaAs[1,2], Si[3,4], InGaAs[5,6], and Ge[7,8]. One of the primary objectives of these studies is to understand spin dynamics[9], i.e. to determine whether the spin polarizations, which can be transported from a ferromagnetic metal through the semiconducting medium, are sustainable over a long enough period of time to be detected with high efficiency. It is known from optical experiments in GaAs[10-12] and electrical experiments in Si[13] that the spin lifetimes of the electrons depend on the carrier density of these semiconducting materials. In the case of GaAs, the optical spin lifetimes of materials which are doped near the metal to insulator transition (MIT) are two orders of magnitude larger than the lifetimes on the metallic and insulating sides[10]. Consequently, it is of great importance to determine the optimal carrier density which will give the longest spin lifetime values for any material which may be used for technological applications.

Here we present electrical spin transport and spin accumulation measurements from Fe electrodes into an $Al_{0.3}Ga_{0.7}As$ channel, which are confirmed via the Hanle effect in the four-terminal (4T) and three-terminal (3T) geometries. Comparison to the spin drift-diffusion (SDD) model for spin transport and the Lorentzian fit for spin accumulation allows one to determine the spin lifetimes in this material. This specific alloy was chosen as the spin transport medium because *it is a persistent photoconductor, which means that the effective carrier density can be tuned in a controlled manner through the MIT using photoexcitation[14]*. This allows one to perform spin-dependent measurements on a lateral spin transport device at various effective carrier densities on one sample *in situ*, potentially enabling a systematic description of the spin lifetime dependence on carrier density and leading to a deeper understanding of how doping levels affect a semiconductor's spin-dependent properties.

There is currently another method which permits the tuning of the carrier density of a material using electrostatic gating[15]. However, utilizing the persistent photoconductivity of $Al_{0.3}Ga_{0.7}As$ in a lateral spin transport device has two noteworthy advantages over gating: 1) gating is generally applicable only to thin layers or quantum wells, but here we can study the spin lifetime dependence in thin as well as bulk AlGaAs, thus enabling direct comparison with the wealth of optical[10,11] and electrical[2] spin lifetime results in bulk GaAs; 2) the electric fields which are present when gating a material will also affect the spin lifetime in the material via the spin-orbit interaction[15], thus making it more difficult to establish its dependence on the carrier density. Our method of increasing the carrier density using photoexcitation will have negligible side-effects on the spin properties in the system.

These spin dependent experiments were performed on MBE-grown Si doped $Al_{0.3}Ga_{0.7}As$ heterostructures with a 2 μm thick active layer (with nominal Si concentration of $1\times10^{19}$ cm$^{-3}$), a 30 nm graded junction of AlGaAs, 5 nm of Fe, and 2 nm of Al which are grown without breaking vacuum. The Al cap layer is grown to prevent the Fe electrodes from oxidizing and the graded junction forms thin Schottky tunnel barrier contacts between the electrode and channel material, which should facilitate efficient spin injection[16]. A lateral spin transport device (Fig. 1(a)) is patterned from this heterostructure using photolithography, wet chemical etching, RF sputtering and thermal evaporation; the structure and dimensions of the device are illustrated in Fig. 1(d).

The device is then wired to a socket for a home-built rotating probe in a JANIS $^4$He cryostat. A commercial infrared LED (λ = 940 nm, E = 1.32 eV) is connected and positioned so that it can shine directly onto the device while in the cryostat (Fig. 1(c)). A commercial GaAs Hall bar is also attached in order to determine the parallel and perpendicular field positions of the rotating probe. The lateral spin transport device is then cooled down from 300 K to 5 K in the dark, during which the persistent photoconductive material will become increasingly insulating (see for instance the black curve in Fig. 1(b), which indicates the resistance versus temperature for a patterned Hall bar of the same AlGaAs). The mechanics of persistent photoconductivity (PPC) are described in a previous paper[17] and an in-depth review is given by Mooney[18]. Essentially, using the photoexcitation from the LED, one can increase the effective carrier density in the material. As can be seen in Fig. 1(b), there is a large decrease in the low temperature resistance (and thus increase in mobile carriers of the system) after illuminating the device for > 24 hours. The effect is termed "persistent" because it remains long after (i.e. weeks) the photoexcitation is discontinued at low temperatures. After the measurements are completed at one specific carrier density, the device can be illuminated again, causing another decrease in resistance and increase in effective carrier concentration. In this way, a set of measurements can be completed at various carrier densities at low temperature on one sample *in situ*.

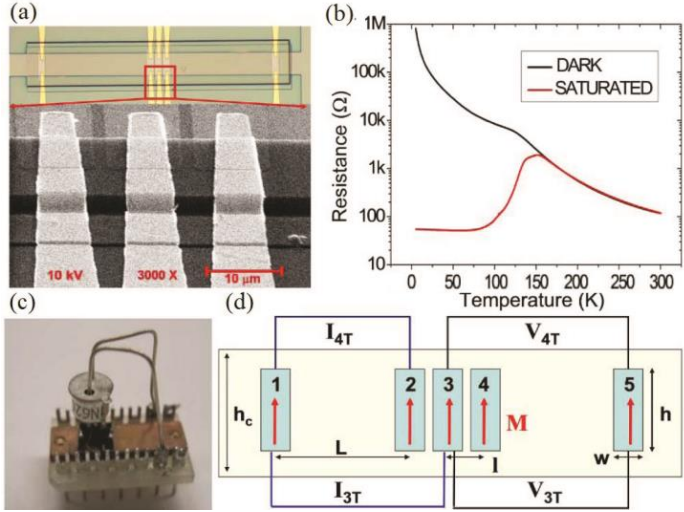

FIG 1: (a) An optical and zoomed-in SEM image of the lithographically defined AlGaAs channel with patterned Fe electrodes, $SiO_2$ isolation pads, and Cr/Au leads. (b) Plots of resistance versus temperature for a Hall bar sample fabricated from the same AlGaAs material as in the spin devices. The black line is the initial cool down in the dark from 300 K to 5 K and the red line is for the same sample under saturated photoexcitation heated from 5 K to 300 K. (c) Photograph of an infrared LED positioned on the socket so it can shine directly onto the sample while inside the cryostat. (d) Schematic (not to scale) of the device and measurement setup. The parameters for this sample are as follows: L = 160 μm, l = 12 μm, h = 50 μm, w = 10 μm, $h_c$ = 70 μm. The Fe bars are 5 nm thick and the AlGaAs channel is 2 μm thick.

The experimental setup is depicted in Fig. 1(d). Before measuring in either the 3T or 4T configuration, the electrodes are first magnetized parallel to each other by applying an in-plane magnetic field. Non-local spin transport measurements (4T) were completed with an AC lock-in technique where the current was applied from electrode 1 to electrode 2. The spins are injected at electrode 2 and diffuse in both directions. A small, varying perpendicular magnetic field is applied, which causes the spins to precess with the Larmor frequency. The spins are detected "non-locally" at contact 3 with no charge current flowing through the detector. For the spin accumulation (3T) measurements, a DC current of 200 µA is applied in all measurements from electrode 1 to electrode 3, which acts as both the injecting and detecting electrode (Fig. 1(d)). The spins which accumulate in the channel underneath contact 3 will also dephase if a perpendicular magnetic field is swept.

The carrier density of the sample can be inferred by comparing the resistivity of the device to that of a Hall bar of the same material. The carrier density, resistivity, and mobility were determined as a function of illumination time for the reference sample (sample S1 as described in detail in Ref. 17). Also, a magnetic characterization of the Fe electrodes has been carried out and the coercive fields were determined for various thicknesses and aspect ratios[19].

Fig. 2 represents the electrical spin injection, accumulation and detection in AlGaAs. These Hanle effect measurements in the 3T geometry measure the accumulation of spins in an AlGaAs channel underneath a ferromagnetic electrode for four different carrier densities which span the MIT (occurring at $9.0 \times 10^{16}$ cm$^{-3}$ at 5 K in this material[17]) using the same bias current of 200 µA. The Hanle signal emerges after the background is subtracted. The signal is determined by first taking the symmetric part of the raw data to remove the anti-symmetric background including the linear contribution likely from the local Hall effects[2] (the largest portion of the raw signal) and then by subtracting an even polynomial from the signal to eliminate the smaller remaining symmetric background. The polynomial is chosen to best-fit the high-field background (beyond the range of the Hanle signal).

Here the data are modeled using a Lorentzian fit as given by:

$$L_z(B) = L_0 / [1 + (g\mu_B B \tau_s / \hbar)^2] \qquad (1)$$

where $L_0$ is the amplitude of the curve[11]. The full width at half maximum (FWHM) then can be determined by setting $L_z$ equal to half of $L_0$ and solving for $B$ which allows one to determine the spin lifetime $\tau_s$ of the material[4]. Since it is possible to overestimate or underestimate the background in the low-field region, leading to a narrowing or broadening of the resulting Lorentzian signal, different background subtractions were tested and the $\tau_s$ value can vary by as much as a factor of two with different acceptable polynomial fits. The spin lifetimes quoted in the inset of Fig. 2 correspond to using the best fit to the background. The same procedure was used to determine the $\tau_s$ for each carrier density, so the trend should not be altered regardless of the background subtraction.

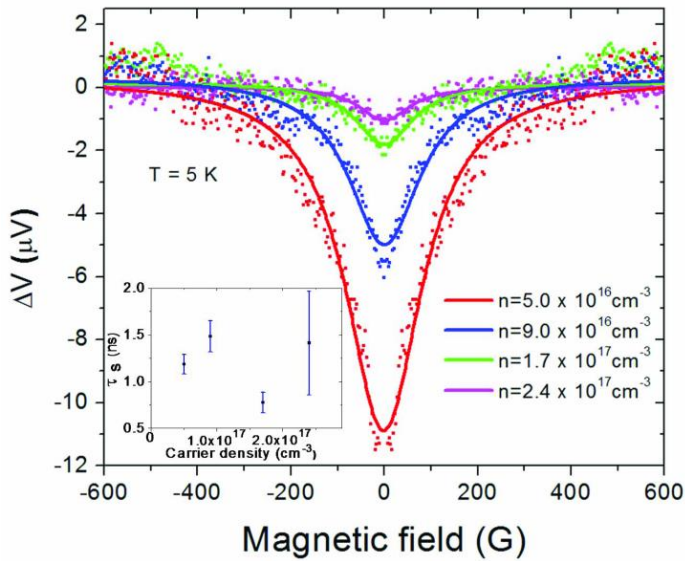

FIG 2: Spin accumulation measured using a DC current in the 3T geometry at 5 K for four different carrier densities at a bias of 200 μA. The data in blue are approximately at the MIT. All curves are modeled via a Lorentzian fit, which yields the spin lifetimes plotted in the inset.

The deduced spin lifetime values are 1.19 ± 0.11 ns, 1.48 ± 0.17 ns, and 0.77 ± 0.11 ns for the carrier density on the insulating side, at the transition (approximately), and on the metallic side respectively (where the errors represent the fitting error of the Lorentzian signal to the best-fit background subtraction). The most metallic datum point has a much larger error since the signal to noise ratio is so low. Using a higher bias would increase the signal-to-noise ratio but decrease the spin lifetime; it would also prohibit one from comparing the lifetimes at different carrier densities[20]. However, this spin accumulation method is clearly adequate for measuring how the spin lifetimes vary with carrier density all at one bias over a broad range across the MIT.

These spin lifetime values can be compared to previous results obtained electrically and optically in GaAs. The electrical spin lifetime of GaAs doped on the slightly metallic side of the transition has been reported as 24 ns at 10 K[2], which is about an order of magnitude less than what is obtained optically in the same material[10,11]. It is natural to assume that the spin lifetime would be even shorter still in AlGaAs, due to the increased disorder of the alloy. In fact, optical measurements of $Al_xGa_{1-x}As$ quantum wells with varying Al concentration $x$ show spin lifetimes between 100 and 450 ps at 5 K[21]. Furthermore, time resolved Faraday rotation measurements in heavily doped bulk AlGaAs have yielded optical spin lifetimes around 2.5 ns in this material[22], which are consistent with our initial electrical results.

It is common for the 3T spin accumulation to be observed in a device even when the non-local 4T spin transport is not detectable. This is because in general it is harder to measure the transport of spins since the signal sizes are smaller and the parameters of the device must be optimized. Next, we report electrical spin injection, transport, and detection in AlGaAs utilizing this 4T configuration. These results are shown in Fig. 3 for two different carrier densities which are both on the insulating side of the MIT. This achievement demonstrates that spins can be transported, manipulated, and detected through even very highly disordered alloy systems. The raw signal obtained from this measurement has an offset of -50 mV, which is several orders of magnitude larger than the spin dependent signal. This offset is due primarily to the current spreading out in the transverse direction[2] of the AlGaAs channel and from unknown background contributions which are not spin-dependent. After subtracting the offset, the background is removed using the same procedure as for the accumulation data in order to determine the Hanle signal. However, in the 4T data, there is a much smaller symmetric contribution to the background.

The resulting Hanle signal is analyzed using the one-dimensional SDD model which incorporates spin relaxation and spin precession in a magnetic field:

$$\frac{\partial \vec{S}}{\partial t} = D\frac{\partial^2 \vec{S}}{\partial x^2} - v_d \frac{\partial \vec{S}}{\partial x} - \frac{\vec{S}}{\tau_s} - \vec{\Omega}_L \times \vec{S} \qquad (2)$$

where $\vec{S}$ is the spin polarization, $v_d$ is the drift velocity, $\tau_s$ is the spin lifetime, $\vec{\Omega}_L = g\mu_B \vec{B}/\hbar$ is the Larmor frequency, $\mu_B$ is the Bohr magneton, $g$ is the electron g-factor, $\vec{B}$ is the magnetic field and $\hbar$ is the reduced Planck's constant. Eq. (2) can be solved using separation of variables and a Fourier analysis, assuming initial conditions and that the spins only precess in the x-y plane. The solution in the x direction is:

$$S_x(x_2, x_1, B, t) = \frac{S_0}{\sqrt{4\pi D t}} e^{-(x_2-x_1-v_d t)^2/4Dt} e^{t/\tau_s} \cos(\Omega_L t) \tag{3}$$

where $S_0$ is the spin injection rate per unit length, $x_1$ is the source of the spins, and $x_2$ is the detection point. This steady state equation can be obtained by integrating over time and the spatial distribution of the contacts. In our case, $v_d = 0$ because there is no current flowing through the detector, $D$ can be determined from the Hall and resistivity measurements on the calibration sample[17], and $g = 0.44$ for $Al_{0.3}Ga_{0.7}As$[23]. The data can be fit using this model with $S_0$ and $\tau_s$ as the only free parameters.

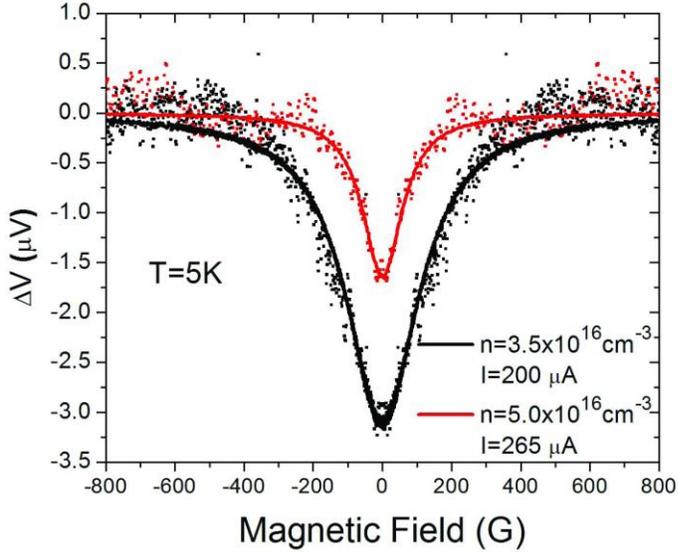

FIG 3: Spin transport Hanle measurement using an AC current in the 4T geometry at 5 K. Both carrier densities are on the insulating side of the MIT. The transport is modeled using the spin drift-diffusion model with spin injection rate $S_0$ and spin lifetime $\tau_s$ as fitting parameters.

The amplitude of the transport signal is smaller than for the accumulation signal, which is consistent with what is found in the literature[13,24-26]. Also, the amplitude of the signal decreases with increasing carrier density, consistent with what was observed in the accumulation data. For a carrier density of $3.5 \times 10^{16}$ cm$^{-3}$ and a bias of $200 \pm 10$ μA, the spin lifetime (as determined by the SDD model) is $2.86 \pm 0.74$ ns. For a carrier density of $5.0 \times 10^{16}$ cm$^{-3}$ and a bias of $265 \pm 8$ μA, the spin lifetime is increased to $4.46 \pm 0.86$ ns. Using the Lorentzian fitting method instead for these data gives spin lifetimes of $1.32 \pm 0.36$ ns and $1.78 \pm 0.33$ ns respectively, about a factor of 2 less. This is consistent because the Lorentzian fitting procedure is recognized to experimentally give a lower bound of the spin lifetime[4]. Further illumination of the device past the MIT ($9.0 \times 10^{16}$ cm$^{-3}$) results in a Hanle transport signal which is difficult to distinguish from the background in the raw data due to decreasing signal to noise.

Through the combination of the 3T and 4T results, we see a consistent picture of the spin lifetime dependence on carrier density emerge. There is an increase in spin lifetime on the insulating side approaching the MIT; on the more metallic side, the spin lifetime is reduced, as one would expect from the Dyakonov-Perel-like relaxation mechanism which is found in GaAs[11,12]. However, in optical GaAs experiments this increase at the MIT is two orders of magnitude larger than on the insulating and metallic sides, and here we see an increase of only about a factor of two. This study suggests that the heavy dependence of spin lifetime on carrier density found in optical experiments in GaAs does not hold for electrical experiments in AlGaAs.

In conclusion, electrical spin transport and accumulation have been achieved in a lateral spin transport device utilizing AlGaAs as the spin transport medium. This demonstrates that the added disorder to the system (Al alloying and high Si doping for the PPC effects) does not preclude electrical measurement of the spin dynamics in this material. However, the electrical spin lifetimes are about an order of magnitude less than those reported for GaAs via electrical measurements. Here, the PPC effects allow for spin dependent properties to be determined at various carrier densities all on one and the same sample via photodoping, without removing it from the cryostat and the need to fabricate many devices.

This work sets a precedent for upcoming spin transport experiments utilizing PPC materials, which will allow for the systematic study and provide a detailed understanding of how doping levels in semiconductors can affect their spin lifetimes. This research may offer useful guidelines for materials optimization for semiconductor spintronics applications.

The authors would like to thank Xiaohua Lou and Paul Crowell for useful discussions and Gian Salis and David Awschalom for providing optical spin lifetime characterization of AlGaAs prior to publication. This work has been supported by NSF under Grant No. DMR-0908625 and NSFC under Grant No. 10920101071.